# Deciphering capacitance frequency technique for performance limiting defect state parameters in energy harvesting perovskites

Vikas Nandal,*[a,b] Sumanshu Agarwal,[c] and Pradeep R. Nair*[b]

[a] Global Zero Emission Research Center, National Institute of Advanced Industrial Science and Technology, Tsukuba 16-1 Onogawa, Tsukuba, Ibaraki 305-8569, Japan.

[b] Department of Electrical Engineering, Indian Institute of Technology Bombay, Mumbai-400076, Maharashtra, India

[c] Department of Electronics and Communication Engineering, Institute of Technical Education and Research, Siksha 'O' Anusandhan University, Bhubaneswar, Odisha—751030, India

*Email: nk.nandal@aist.go.jp, prnair@ee.iitb.ac.in

*Abstract* — With emerging thin film PIN based optoelectronics devices, a significant research thrust is focused on the passivation of trap states for performance enhancement. Among various methods, capacitance frequency technique (CFT) is often employed to quantify trap state parameters, however, the trapped charge induced electrostatic effect on the same is not yet established for such devices. Herein, we present a theoretical methodology to incorporate such effects in the CF characteristics of well-established carrier selective perovskite-based PIN devices. We show that the electrostatic effect of trapped charges leads to non-linear energy bands in perovskite layer which results in the underestimation of trap density from existing models of CFT. Consequently, a parabolic band approximation with effective length PBAEL model is developed which accurately predicts the trap density for shallow or deep states from CFT analysis. In addition, we demonstrate that the attempt to escape frequency, crucial for trapped charge dynamics with continuum energy bands, can be well extracted by eliminating non-linear effects at reduced perovskite thickness. We believe that our work provides a unified theoretical platform for CFT to extract trap state parameters for a broad class of organic and hybrid materials-based thin film devices for energy conversion applications such as solar cells, LEDs, etc.

## 1 Introduction

With excellent optical properties, low-cost solution processed organic and hybrid materials have attracted a significant research interest towards the development of optoelectronics devices such as solar cells, LEDs, etc.[1–6] The performance of various energy conversion devices is significantly limited by shallow and/or deep defect states.[7–11] For instance, free charge carriers in respective continuum energy bands undergo trapping/detrapping process with shallow defect states which leads to reduced charge carrier mobility. In contrast, deep trap states act as Shockley-Read-Hall (SRH) recombination centres and increase the recombination loss of photogenerated charge carriers.[12–15] Therefore, material characterization is imperative to quantify such performance limiting trap state parameters like trap density, trap energy, and attempt to escape frequency.

In literature, depending on device architecture and experimental conditions, various measurements methods like capacitance frequency technique (CFT),[16–18] deep level transient spectroscopy (DLTS),[17,19,20] thermally simulated current (TSC),[21–23] capacitance voltage (CV),[24] and thermally dielectric relaxation current (TDRC) are employed to obtain above mentioned trap state parameters.[25–27] In particular, CFT is a well-established and widely utilized characterization scheme for emerging organic, inorganic and hybrid materials e.g. perovskites.[28–32] Using CFT, Walter et al. proposed a linear band approximation (LBA) model $N_l(E)$ for homogenous PIN diode and parabolic band approximation (PBA) model $N_p(E)$ for PN diode to estimate trap density of semiconducting material inside depletion region and were approximately given by[16]

$$N_l(E) \approx -\frac{V_{bi}^2}{qALkT}f\frac{dC}{df}, \quad (1)$$

$$N_p(E) \approx -\frac{V_{bi}}{qALkT}f\frac{dC}{df}. \quad (2)$$

Here, $E$ is energy with respect to continuum energy bands, $V_{bi}$, $L$, $A$, $T$, and $C$ are the built-in voltage, depletion width, cross-sectional area, lattice temperature, and capacitance of the diode, respectively. The constants $q$, $k$, and $f$ are the electronic charge, Boltzmann's constant, and frequency of superimposed AC modulating signal for capacitance measurement. As the name suggest, these models were strictly based on the assumption that the depletion region of the diodes exhibited either linear or parabolic energy band diagram. However, in practical thin film solar cell or LED devices, the density of trap states could be significantly large which may impart additional charging or electrostatic effect on the energy bands in the depletion region. Such electrostatic effect of trapped charges in the depletion region was not considered in the above-mentioned models which could lead to erroneous estimates of trap states parameters and attempt to escape frequency from CFT. To the best of our knowledge, the applicability of CFT to carrier selective PIN based systems, where the current is determined by recombination rate of charge carriers, is not yet theoretically explored in the literature. All these indicate that a modified methodology is needed to reliably estimate parameters associated with traps in emerging device concepts using perovskite materials.

In this work, we introduce a methodology to include the electrostatics effect in CFT for carrier selective PIN based devices. The methodology is validated using a widely explored carrier selective perovskite-based PIN device, with trap/defect states as shown in Figure 1. Our results indicate that the electrostatic redistribution originated from the trapped charges in the intrinsic perovskite layer leads to non-linear energy bands. As a result, the existing LBA and PBA model underestimate the trap density of perovskite. Consequently, we develop a model for non-linear bands which provides an excellent estimate of (shallow or deep) trap states from CFT. In addition, we show that the extraction of attempt to escape frequency from CFT depends on device electrostatics and can be accurately measured by decreasing the thickness of perovskite material. Next, we present the details of model system and simulation strategies for CFT analysis.

## 2 Model system

Figure 1(a) shows the schematic of planar configuration of perovskite-based PIN diode. CH$_3$NH$_3$PbI$_3$, with band gap $E_g = 1.55\ eV$ and thickness $L = 400\ nm$, is the intrinsic perovskite layer which is sandwiched between electron transport layer (ETL) and hole transport layer (HTL). The transport layers are chosen in such a way that the alignment of conduction and valence energy bands with perovskite layer ensure a selective transport of the charge carriers in the device. For instance, a heavily doped n-type electron transport layer (ETL) and a p-type hole transport layer (HTL) injects only electrons and holes into perovskite layer, respectively. Here, the dark current is mainly governed by the recombination of injected charges carriers inside perovskite layer.[33] Reports indicates that perovskite comprised of significantly high density of fixed trap states depending on the fabrication method.[27-31] Consequently, we introduce the fixed acceptor type trap states in the forbidden gap of perovskite layer. Because of lack of information on the spatial distribution of trap density in perovskite, we consider that the defect states are uniformly distributed across the thickness of perovskite layer. Moreover, the assumption is helpful to model the problem in a computationally tractable manner. The distribution of trap density $N_i(E)$ in energy $E$ shown in Figure 1(b) is modelled by Gaussian function, and is given as

$$N_i(E) = N_t \exp\left(-\frac{(E-E_t)^2}{2\sigma^2}\right). \quad (3)$$

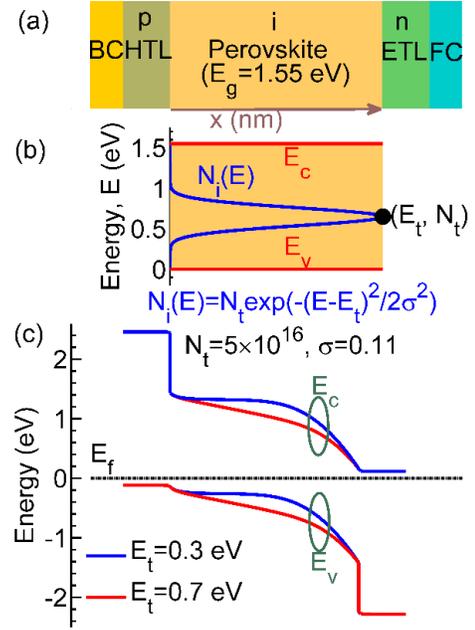

**Figure 1. Effect of trapped charges on the electrostatics of the perovskite-based PIN device.** (a) Schematice of the modeled device incorptorated with (b) gaussian distribution of acceptor type trap density $N_i(E)$ with maximum trap density $N_t$ at energy $E_t$ inside energy band gap of perovskite layer. (c) Equilibrium band diagram for different trap energy $E_t$. Trapped charges induces non-linear (conduction $E_c$ and valence $E_v$) energy bands inside perovskite layer which depends on the $E_t$.

Here, $N_t\ (cm^{-3}eV^{-1})$ is the maximum trap density of states present at trap energy $E_t\ (eV)$, $\sigma$ is the spreading or disorder factor in energy $(eV)$. The details of the material parameters employed for numerical simulations are provided in Table S1 of supplementary Information (SI). In particular, the employed values of $N_t$, $\sigma$ and attempt to escape frequency in numerical model are similar or close to the experimental data reported in the literature,[28] whereas the trap energy $E_t$ is varied over wide energy range to model various shallow or deep trap states.

Using TCAD Sentaurus device simulator tool,[34] detailed numerical simulations of the modeled PIN diode are performed by solving drift-diffusion, Poisson's and continuity equation for electrons and holes self-consistently.[35] Figure 1(c) shows the effect of trapped charges on electrostatics of perovskite-based PIN diode. In the absence of trapped charges, the energy bands vary linearly along the thickness of intrinsic perovskite layer.[36–38] However, the presence of trapped electrons in acceptor type trap states (present below fermi level) imparts immobile negative charges, whereas, empty trap states (above fermi level) do not provide any charge effect on the device electrostatics. This results in non-linear energy bands in the perovskite layer. Such electrostatics redistribution depends on trap density $N_t$, trap energy $E_t$, and dielectric constant of perovskite layer.

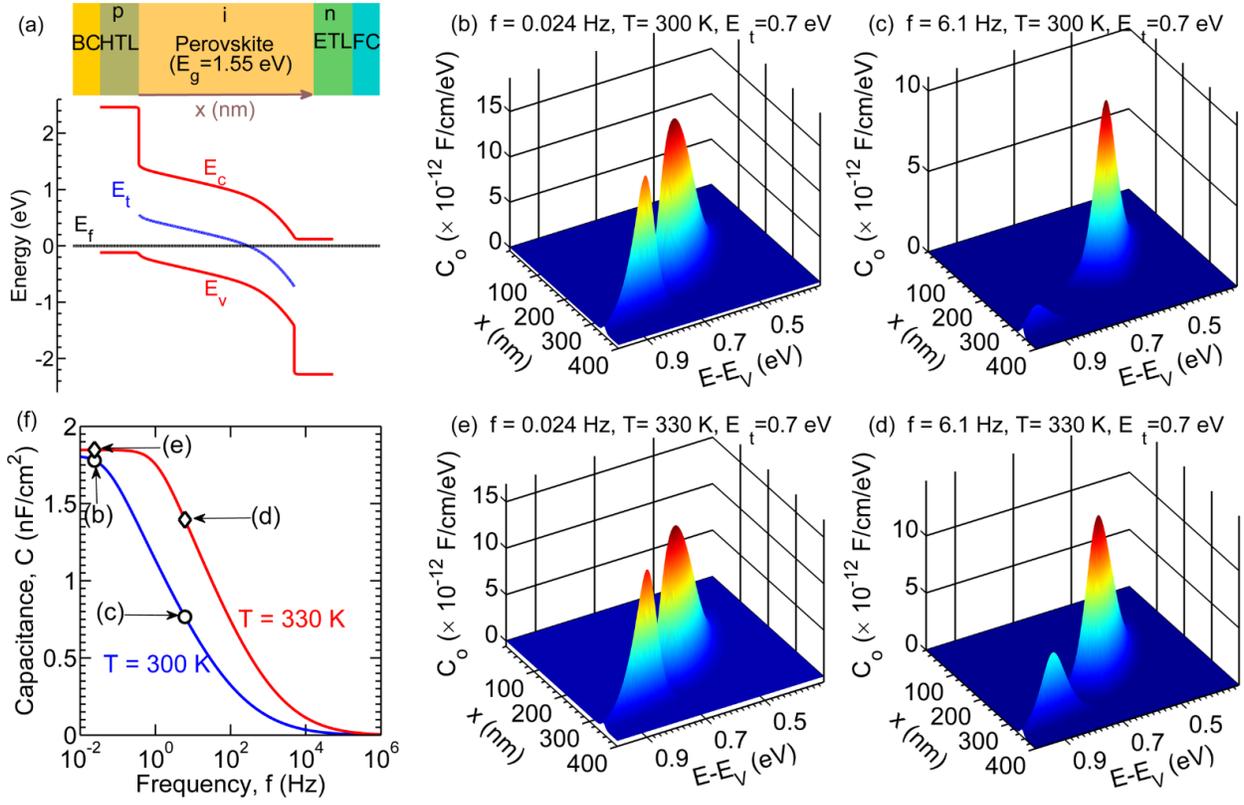

**Figure 2. Impact of frequency f and temperature T on the capacitance of perovskite-based PIN device.** (a) Device schematic and equilibrium band diagram, where the trap states (in blue) are present at trap energy $E_t = 0.7\ eV$ (w.r.t. valence band). (b-e) Mapping of calculated capacitance $C_o$ over position $x$ and energy $E$ inside perovskite layer at frequency $f$ and temperature $T$. (f) capacitance frequency characteristics at temperature $T = 300$ and $330\ K$. The integrated values of capacitance $C_o$, shown in (b)-(e), are represented by the associated symbols. The capacitance reduces with the increase in frequency $f$, however, such decrease in capacitance is recovered with the increase in $T$.

## 3 Results & discussion

To elucidate the shortcomings of the existing models for estimating the trap density in PIN diodes, we perform capacitance vs frequency (CF) simulations at various temperature $T$ for perovskite-based PIN diodes in the presence of the traps and the results are used to back extract the trap state parameters. For this, at applied voltage $V = 0$, we employ simulated (using TCAD) variation of energy band along with electron density $n$ and hole density $p$ with position $x$ for the calculation of capacitance. As reported by Walter et al.,[16] the capacitance $C_o$ is originated from the modulation of trapped charges with the superimposed AC signal amplitude $u$ at frequency $f$ and is given by

$$C_o = \frac{q}{kTu}\frac{f_1}{f^2 + f_o^2}\big(-\beta_n n q u_n + f_n(E)(\beta_n n q u_n + \beta_p p q u_p)\big)N_i(E). \quad (4)$$

Here, $\beta_{n,p}$ is carrier capture coefficient, $f_n(E)$ is the fermi function, $qu_{n,p}$ is the modulation of quasi-fermi levels with AC signal. Subscripts $n$ and $p$ correspond to electrons and holes, respectively. Futher, $f_1$ and $f_o$ are given as

$$f_1 = \beta_p p\left(1 + \exp\left(-\frac{E-E_f}{kT}\right)\right) - \beta_n n\left(1 + \exp\left(-\frac{E_f-E}{kT}\right)\right), \quad(5)$$

$$f_o = \beta_p p\left(1 + \exp\left(-\frac{E-E_f}{kT}\right)\right) + \beta_n n\left(1 + \exp\left(-\frac{E_f-E}{kT}\right)\right), \quad(6)$$

where, $E_f$ is the fermi-energy level. As evident from Eqs. 4-6, the capacitance $C_o$ depends on position $x$ inside perovskite layer and energy $E$. Further, this method allows the estimation of $C_o$ as a function of temperature $T$ and frequency $f$ which will be later used to extract the trap state associated parameters. The total capacitance $C$ is calculated by numerical integration of $C_o$ along the position $x$ and energy $E$ i.e., $C = \iint C_o dx dE$. Above mathematical model formulation in Eq. 4 provides the capacitance originated from the fixed trap states and does not consider the contribution from ion migration. Incorporation of ion migration induced capacitive effect is non-trivial, and therefore, currently out of the scope of the current manuscript. Futscher *et al.* reported that the distinction between electronics trap states and mobile ions requires frequency and time domain analysis of capacitance signals.[39] Despite similar capacitance signatures, transient capacitance measurements indicated that the ratio of capacitance rise time to the decay time is greater and less than 1 for mobile ions and electronic trap states, respectively. Such contrasting transient characteristics were attributed to the different time scales of underlying physical processes of trapping and detrapping for trap states, whereas, diffusion and drift times for mobile ions.

Figure 2a presents the schematic of the device under investigation along with the equilibrium energy bands. The trap states with trap energy $E_t = 0.7\ eV$ introduces a non-linear band bending in the perovskite layer. The calculated capacitance $C_o$, using simulation results and Eqs. 4–6, is mapped over position $x$ and energy $E$ at different frequency $f$ and temperature $T$ in Figure 2(b-e). By definition, here, the capacitance is a measure of rate of trapped charge modulation with the superimposed applied AC voltage. Such charge modulation is limited by the dynamic response of trapped charges with contimuum energy bands which is governed by emission rate $e$ as[16]

$$e = 2\beta_{p,n} N_{c,v} \exp\left(-\frac{E_a}{kT}\right). \quad (7)$$

Here, $N_{c,v}$ is the effective density of states in conduction $E_c$/valence band $E_v$, $E_a$ is the activation energy or energy depth of trap state from $E_c$ or $E_v$, and $\beta_p$ ($\beta_n$) is the product of capture cross sectional area $c_p(c_n)$ of trap states and average thermal velocity $v_{th,p}(v_{th,n})$ of charge carriers. The prefactor $2\beta_{p,n} N_{c,v}$ is defined as attempt to escape frequency $f_{AEF}$. From Eq. 7, for a fixed $f_{AEF}$, it is evident that the dynamic response of trapped charges contributing towards capacitance decreases with increase in energy depth E, whereas, the same increases with the rise in temperature $T$. Accordingly, Figures. 2(b, c) show the impact of frequency $f$ on capacitance $C_0$ which indicate that the rise in $f$ from 0.024 to 6.1 Hz leads to the decrease in $C_0$. We find that such decrease in $C_0$ is relatively higher at high energy $E$ than low energy which is originated from the assymetric response of trapped charges present at different energy, as per Eq. 7. For instance, the response of trapped charges in deep trap states is less than shallow states (with respect to valence band) of $N_i(E)$. Similar behavior of the $C_0$ decrease with $f$ is also observed at a higher temperature in Figs. 2(d, e). Besides, at a relatively high value of $f = 6.1\ Hz$ in Figures 2(c, d), the $C_0$ decrease is partially compensated with an increase in temperature from $T = 300$ to $330\ K$. Since the total capcitance $C$ is obtained by the numerical integration of $C_0$, similar effects of frenquency $f$ and temperature $T$ are translated in CF characteristics, presented in Figure 2(f).

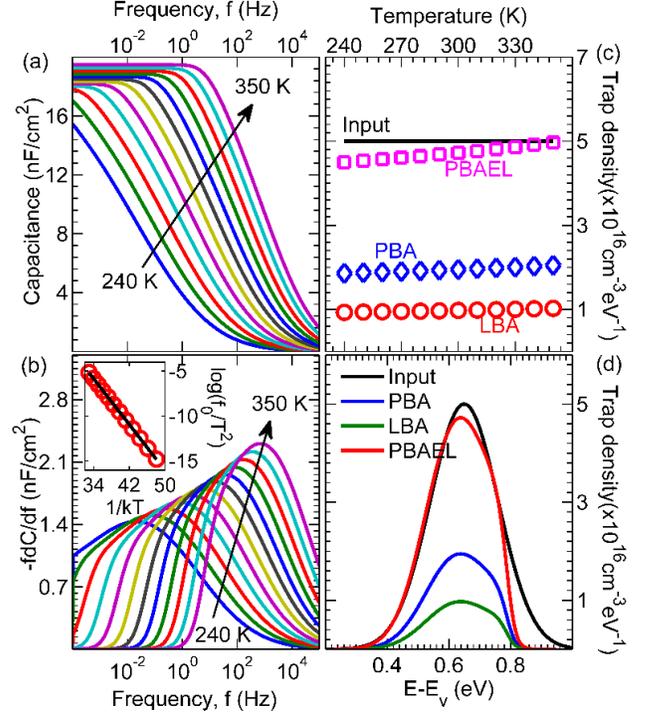

**Figure 3. Estimation of trap states parameters in perovskite-based PIN device**. a) Capacitance frequency (CF) characteristics at various temperature ($240\ K - 350\ K$ in steps of $10\ K$), b) CF curves plotted as $-fdC/df$ vs. $f$ to extract frequency $f_0$ at temperature $T$. Inset shows $\log(f_0/T^2)$ versus $1/kT$ from CF curves (symbols) which is fitted with Eq. 7 (solid line). c) Estimated (c) trap density and (d) its energy distribution obtained from CF analysis using existing models (parabolic band approximation-PBA $N_p$; linear band approximation-LBA $N_l$) and developed model (Parabolic band approximation with effective length-PBAEL $N_{PEL}$) along with the actual input employed in modeled device. The results show that the existing models underestimate the trap density and its energy distribution, whereas our model provides good estimates in comparison to the actual input.

In Figure 3, the CF characteristics over wide range of temperature from $T = 240$ to $350\ K$ are further computed and analysed to estimate trap state parameters. Figure 3(a) indicate that the capacitance remains constant up to a certain threshold frequency as all trapped charges are able to respond across $N_i(E)$ respond to the modulating signal. Beyond threshold frequency, the asymetric lagging response of trapped charges (shown in Figure 2) leads to the decrease of capacitance with frequency. We observe that the onset or threshold frequency for such decrease in capacitance increases with the increase in temperature due to improved emission rate with the temperature rise as per Eq. 7. From Eq. 1 and 2, the trap density is proportional to the negative of the product of frequency $f$ and rate of change of capacitance $C$ with $f$. Therefore, we obtain $-fdC/df$ vs. $f$ characteristics at different temperatures $T$ in Figure 3(b). The results indicate that these curves exhibit increasing and decreasing trends with the frequency at various $T$. For a particular $T$, such dual trend leads to $-fdC/df$ maximum at a defined frequency $f_0$ which is attributed to the emission rate of trapped charges from trap energy $E_t$ such that $f_0 = e = f_{AEF} \exp\left(-\frac{E_t}{kT}\right)$. As a result, the frequency $f_0$ increases with the increase in temperature T. Considering $f_{AEF} \propto T^2$ (ref. [35]) the $\log(f_0/T^2)$ is plotted against $1/kT$ in inset of Figure 3(b) to back extract trap energy $E_t$ and $f_{AEF}$ from the slope and point of intersection with y axis of the linear fit, respectively. The estimated trap energy $E_t = 0.638\ eV$ and $f_{AEF} = 6.9 \times 10^{11}\ Hz$ which are in reasonable agreement with the input and experimental values of $0.65\ eV$ and $3 \times 10^{11}\ Hz$, respectively.[28]

Besides $E_t$ and $f_{AEF}$, the trap distribution profile $N_i(E)$ is obtained from $-fdC/df$ vs. $f$ characteristics by the (a) conversion of frequency $f$ axis to energy $E$ axis, and (b) scaling of $-fdC/df$ to trap density $N_i(E)$. By rearranging Eq. 7 with $E - E_v = E_a$ and $f = e$, the frequency $f$ is mapped to energy depth $E - E_v$ by using

$$E - E_v = kT \log\left(\frac{f}{f_{AEF}}\right), \quad (8)$$

whereas the scaling factor (b) depends on the nature of band profile in the depletion region. Since the nature of band profile is not accessible through experiments, we examine both existing LBA (in Eq. 1) and PBA (in Eq. 2) models and find a suitable model for trap distribution profile $N_i(E)$ by comparing the estimated profile from CF analysis with the input. Figures 3(c) and 3(d) show the back extracted trap density $N_t$ at different temperature and distribution of trap density $N_i(E)$ at temperature $T = 300\ K$, respectively. The results display that the estimated values of $N_t$ are independent of temperature $T$. In addition, the existing models underestimate the trap density $N_t$ and hence $N_i(E)$. We believe that such mismatch between extracted trap density from CF analysis and the input is primarily arises from non-linear energy bands of perovskite layer (shown in Figure 2(a)) which is neither linear nor parabolic. Therefore, for a non-linear energy bands, we introduce a parabolic band approximation model incorporated with effective length (PBAEL) $N_{PEL}$ to incorporate the electrostatic effects of trapped charges and is approximately given as

$$N_{PEL} \approx -\frac{V_{bi}}{qAL_{eff}kT} f \frac{dC}{df}. \quad (9)$$

Here, $L_{eff}$ is the effective length or space charge width created by electrostatics effect of trapped negative charges inside perovskite layer. Compared to PBA model $N_p$, the physical length of intrinsic layer reduces to the effective length of depletion region. For $V_{bi}$ and $L_{eff}$, numerical simulations (using TCAD simulator) were performed to obtain capacitance voltage (CV) characteristics in Fig. S1 of SI. The effective length or space charge width $L_{eff} = \epsilon A/C$ ($\epsilon$ is the electrical permittivity of perovskite) is obtained at $V = 0$, whereas $V_{bi}$ is extracted through Mott-Schottky analysis of CV characteristics. From Figures S1(a) and S1(b) of SI, we find that the extent of parabolic region in perovskite layer increases with the increase of bias voltage in reverse bias regime. With estimated $V_{bi}$ and $L_{eff}$ from Fig. S1 of SI, we find that our PBAEL model $N_{PEL}$ (given by Eq. 9) provides excellent estimates of trap density profile (as compared to input and reported data)[28] over a wide range of temperature $T$. The robustness of and flexibility of our model for shallow and deep trap states is tested by analysing CF characteristics with various trap energy from $E_t$=0.3 eV to 0.65 eV and $N_t = 5 \times 10^{16}\ cm^{-3}eV^{-1}$. As evident from Figure S2 of SI, our model indeed provides good estimates of trap density of states at different trap energy in comparison to existing models.

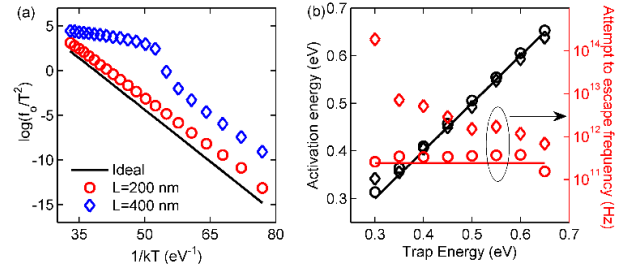

**Figure 4. Electrostatic effect of trapped charges on attempt to escape frequency in perovskite-based PIN device.** (a) Arrhenius plot, $\log(f_0/T^2)$ vs. $1/kT$, extracted from CF analysis (in symbols) for trap energy of 0.3 eV and trap density of $5 \times 10^{16}\ cm^{-3}eV^{-1}$ at different perovskite thickness $L$. Here, solid line represents to an ideal condition of linear energy band in perovskite layer (without trapped charge induced electrostatic effect on energy bands). b) Impact of trap energy and thickness $L$ (200 nm – circles; 400 nm – diamonds) on the back extracted activation energy (left axis) and attempt to escape frequency (right axis) at 300 K. Here, solid lines represent to the respective actual input values.

As discussed earlier, $f_{AEF}$ plays an important role in determining the dynamics of charge carriers between trap states and delocalized states. We observe that the back extraction of $f_{AEF}$ is very sensitive on the Arrhenius plot (of $\log(f_0/T^2)$ vs. $1/kT$) obtained from CF analysis which we believe depends on the electrostatics of perovskite layer. To validate this, the electrostatics is varied by changing trap energy and perovskite thickness $L$ and observe its effect on the Arrhenius plot. In Figure 4, ideal characteristics represent the linear band profile which would result in accurate interpretation of trap energy and attempt to escape frequency $f_{AEF}$. Specifically, Figure 4(a) shows that the Arrhenius plot for $L = 200\ nm$ is close to the ideal case and thereby provides a good estimate of attempt to escape frequency for trap energy $E_t = 0.3\ eV$. The results indicate a significant mismatch between Arrhenius plot for $L = 400\ nm$ and the ideal case. As a result, in Figure 4(b), the $f_{AEF}$ is overestimated by nearly three orders of magnitude than the actual value for trap energy of $0.3\ eV$ for $L = 400\ nm$. Interestingly, the results display that such an error in $f_{AEF}$ decreases with the increase in trap energy $E_t$. In contrast, for $L = 200\ nm$, the estimated $f_{AEF}$ is in reasonable agreement with the actual value for various trap energies and experimental data.[28] These results suggest that the correct estimation of $f_{AEF}$ for shallow and deep trap states requires reduced thickness of perovskite layer thickness to mitigate the electrostatic effect of trapped charges on band bending, whereas the activation energy which corresponds to the trap energy is independent of perovskite layer thickness.

## 4 Conclusions

In sumary, we investigated the electrostatic effect of trapped charges on the back extracted trap state parameters from CF analysis of perovskite based PIN diode. We demonstrate that the trapped charges in perovskite layer modulate linear energy band profile to non-linear band bending in the intrinsic perovskite layer. Such deviation in energy band profile leads to the underestimation of trap density using existing LBA and PBA models. Accordingly, we developed a PBAEL model $N_{PEL}$ which incorporates the electrostatic effect of trapped charges. Our model predicts good estimates of trap denisty for wide range of lattice temperatures and trap energies. In addition, ambiguity in attempt to escape frequnecy $f_{AEF}$ is investigated under different electrostatic conditions of active layer for shallow and deep trap states which suggest that electrostatics played a significant role for the estimation of $f_{AEF}$. The results shows that accurate extraction of $f_{AEF}$ can be obtained by reducing perovskite layer thickness. We believe that our work provides a basic theoretical platform to understand physical insights behind trapped charge induced capacitance to extract trap state paramters for broad range of PIN based solar cells, LED's etc.

## Conflicts of interest

There are no conflicts of interest to declare.

## Acknowledgement

This paper is based upon work supported in part by the Solar Energy Research Institute for India and the United States (SERIIUS), funded jointly by the U.S. Department of Energy (under Subcontract DE-AC36-08GO28308) and the Govt. of India's Department of Science and Technology (under Subcontract IUSSTF/JCERDC-SERIIUS/2012).